\documentclass[aps,pre,floatfix,nofootinbib,showpacs,showkeys,twocolumn]{revtex4-1}
\pdfoutput=1
\usepackage{graphicx}
\usepackage{amsmath}
\usepackage{amssymb}

\begin{document} 
\author{Ramandeep S. Johal}
\email[Corresponding Author: ]{rsjohal@iisermohali.ac.in}
\affiliation{Department of Physical Sciences, Indian Institute of
 Science Education and Research Mohali,
Sector 81, S.A.S. Nagar, Manauli P.O., Mohali 140306, India}
\author{Renuka Rai} 
\affiliation{Department of Applied Sciences,  University Institute of
Engineering and Technology, Panjab University, Chandigarh-160014, India.  }
\author{ G\"{u}nter Mahler}
\affiliation{Universit\"{a}t Stuttgart, 1. Institut f\"{u}r Theoretische Physik, 
Pfaffenwaldring 57 // IV, 70550 Stuttgart, Germany}
\draft
\title{Bounds on Thermal Efficiency from Inference }
\begin{abstract}
The problem of inference is applied to 
the process of work extraction from two constant heat capacity reservoirs,
when the thermodynamic coordinates of the process are not 
fully specified. The information 
that is lacking, includes both the specific value of 
a temperature as well as the label
of the reservoir to which it is assigned. The estimates for 
thermal efficiency reveal that  
uncertainty regarding the exact labels, reduces the maximal efficiency
below the Carnot value, its minimum value being the 
well known Curzon-Ahlborn value. We also make an average estimate 
of the efficiency {\it before} the value of the temperature is revealed.
It is found that if the labels are known
with certainty, then  in the near-equilibrium limit the efficiency
scales as 1/2 of Carnot value, while if there is maximal uncertainty in 
the labels, then the average estimate for efficiency 
drops to 1/3 of Carnot value. We also suggest 
 how infered properties of the incomplete model can be mapped to a 
model with complete information but with an additional source 
of thermodynamic irreversibility.
\end{abstract}
\maketitle
%
\newcommand{\be}{\begin{equation}}
\newcommand{\bea}{\begin{eqnarray}}
\newcommand{\bc}{\begin{center}}            
\newcommand{\ee}{\end{equation}}
\newcommand{\eea}{\end{eqnarray}}
\newcommand{\ec}{\end{center}}
\newcommand{\baa}{\begin{eqnarray*}}
\newcommand{\eaa}{\end{eqnarray*}}
\section{Introduction}
In many situations, we have to reason from incomplete information.
Scientific inference refers to application of a consistent set of principles  
in these situations, which satisfy our rationality. In its initial 
stages, it was termed as the ``the art of conjecturing'' by Bernoulli \cite{Bernoulli2006}. Later 
refined into a technical tool, Laplace \cite{Laplace1774} made a successful use of 
what is now known as Bayes' formula \cite{Bayes1763}.
Later, Cox showed that the only set of consistent axioms justifying plausible reasoning
were the already established axioms of the probability theory \cite{Cox1946, Cox}. 
Many authors have clarified the scope and meaning of inference
\cite{Jeffreys1931,Jeffreys1939,Polya1954,Jaynes2003}.
Following Jaynes, much of the development in thermodynamics 
and statistical mechanics may be regarded as an application of
 the principles of plausible reasoning. 

Usually, the physical mechanisms which generate thermodynamic irreversibility are 
ascribed to a finite rate of heat transfer, internal friction, the finite size
of the reservoirs and so on. However, from the perspective of 
information theory, this irreversibility is also related
to a loss of information about the system into the environment.
In this paper, we consider reversible thermodynamic models but with
incomplete information. 
 The issue we address here is that 
from inference performed on such models, the estimated behavior
exhibits some features of irreversibility. 
Thus in our case, irreversibility does not appear
at the objective physical level, but only in the 
estimated behavior.   
The central feature of our approach is that the missing part 
of information is interpreted in a 
subjective manner, or in other words 
as the observer's lack of knowledge about the system.
Due to lack of complete information, the observer
has to perform inference to estimate the characteristic quantities of the system.
We show that a consistent use of prior information in reversible models 
leads to an estimate for maximal efficiency which is lower than the Carnot value.  
We also suggest that the resulting inferred behavior is analogous to that obtained 
by incorporating explicitly some thermodynamic irreversibility in the 
actual (reversible) process.

The motivation for our approach comes from the connection between
thermodynamics and information. 
This has prompted a fruitful discussion on the role of information
theory in thermodynamic frameworks, for example the
role of Maxwell's demon in information processing \cite{Leffbook}
which is continuing to this day. 
Recently, it has also been explored in
\cite{Johal2010,GRD2012,PRD2012,GPRD2012}, 
that the identification and inclusion of  
prior information in heat cycles with incomplete
specification, leads to interesting analogies with
irreversible models. In particular, many different
 efficiencies show up in the inference based approach
 which are found in the context of time-dependent cycles or
  with other sources of irreversibility \cite{Refeff}. 

Now the prior information,
which is to be exploited in making  inference, can be 
present in different forms, some even quite raw or qualitative. 
 In this paper, 
  we want to understand further how different kinds of prior information,
  impact our expectations about the performance of these heat cycles.
  In particular, we assume an uncertainty not only in the value of a parameter,
  but also an uncertainty in the exact subsystem to which it is assigned.
  The later kind of uncertainty will be addressed as {\it label uncertainty}.
   For example, a classical system may be specified by a set of 
quantities $\{X,Y,..\}$. In the case of a  
multi-partite system, we also label the 
subsystems, say with index $i$. Then the  
properties of all constituent subsystems are distinguished
if our labels are refined as $\{X_i, Y_i,... \}$. 
In the following, we consider a situation 
where the values of the individual parameters ($X, Y,..$) are
known, but we may be uncertain about the exact subsystem
labels. The question is how do we estimate the performance of the system
 based on this incomplete information. 
For simplicity, we will consider only a bipartite set up.

The paper is organised as follows. 
Section II discusses the model and has three subsections.
In subsection A, we introduce a reversible thermodynamic
model of work extraction, with complete specification of all parameters.
In subsection B, we assume uncertainty in the final temperatures
as well as the label uncertainty. The work performed and the
thermal efficiency are estimated for a given measure of uncertainty.
In subsection C, we draw analogy of the infered behavior with 
an explicitly irreversible model of heat engine. Finally, in Section III
we define an average estimate of efficiency, which is calculated 
using a uniform prior distribution over the uncertain temperature. The 
behavior of this average value for near-equilibrium is evaluated,
which leads to establishing two distinct classes for the expected
efficiency, based on zero or complete label uncertainty. The last section IV 
presents the conclusions.  
\section{The Model}
\subsection{The case of complete information}
It is sufficient to consider a textbook example of 
two {\it finite} ideal gas systems with a constant heat capacity $C$,
at initial temperatures $T_+$ and $T_-$ ($T_+ > T_-$), serving 
as the heat source and the sink, respectively.
They are coupled via a reversible work source,
which by design  extracts maximal work due to the available temperature gradient.
At some stage in this process, the initially hot reservoir obtains a 
temperature $T_1$ and the initially cold reservoir
is at temperature $T_2$. The amount of heat
taken in by the engine and the heat rejected
to the sink, are respectively given by $Q_{\rm in} = C (T_{+}-T_{1})$, and
$Q_{\rm out} = C (T_{2}-T_{-})$, respectively.
The total entropy change in the two reservoirs being zero, we have 
$\bigtriangleup S = C \ln {(T_{1}/T_{+})} + C\ln {(T_{2}/T_{-})} = 0$.
This yields
\be
T_1 = \frac{T_+ T_-}{T_2},
\label{a}
\ee
as the relation between the final reservoir temperatures. 
The work performed, $W= Q_{\rm in}-Q_{\rm out}$ is:
$W = C (T_{+}+T_{-}-T_{1}-T_{2})$.
In the following, we set $C=1$.
Finally, using Eq. (\ref{a}), the efficiency $\eta=W/Q_{\rm in}$,
 can be written as:
\bea
\eta &=&  1 - \frac{T_2}{T_+}, \label{et2}\\
     &=&  1 - \frac{T_-}{T_1}. \label{et1}
\eea
We note that the maximum work is obtained if the final temperatures 
obtained are: $T_1 = T_2 = \sqrt{T_+T_-}$, and the efficiency at this 
optimal process is  $\eta = \eta^*= 1-\sqrt{{T_-}/{T_+}}$.

 Now in the standard analysis, $T_+$ and $T_-$ are the fixed 
initial values of the temperatures, 
and due to relation (\ref{a}),
we may regard all the expressions as functions
of only one of the two temperatures, $T_1$ or $T_2$.
 Thus the work performed, can be rewritten as
\be
W(T_2) = \left( T_+ +T_- -T_2  -\frac{T_+ T_-}{T_2} \right),
\label{wt2}
\ee 
with a similar expression in terms of $T_1$.

Now note that  just from the work expression, Eq. (\ref{wt2}),
it is not obvious as to which temperature
is chosen as the variable. We have to look
at the expression for the heat exchanged to assess 
the label corresponding to a specific reservoir.  
So when we have an exact knowledge about the temperatures,
this information has two parts: i) the individual
values of the temperatures and ii) the labels for the 
reservoirs to which a value is assigned. Thus the symbol $T_2$ denotes 
the temperature value of the particular reservoir (label 2). 
\subsection{Incomplete information}
Now let us imagine  a controller
of the process who knows the final thermodynamic coordinates,
or the temperatures of the reservoirs. 
The controller invites us to
play a game of guessing and promises to reveal  one of the values of
the temperatures, but {\it not}
the reservoir to which this value belongs.
The task ahead of us is to infer 
the performance of the engine by making estimates about 
work performed, efficiency and so on.

As mentioned above, the work expression does not reveal 
unambiguously the individual labels of the reservoirs.
Thus given some temperature value $T$, the work expression
(written devoid of reservoir labels) will be 
$W(T) =  T_+ +T_- -T  -{T_+ T_-}/{T}$.
Next comes the issue of the range of possible values
for the final temperature $T$. This should be fixed from the
information that is actually available. In particular, 
we invoke the fact that the extracted work satisfies: $W \ge 0$,
so that $T$ is allowed to take values in the range $[T_-,T_+]$.
Thus to some extent, we  have removed the problem from
its physical context which involves such notions as
the flow of heat from a hot to cold temperature and so on.
In the spirit of an inference based approach, 
we seek to quantify
our beliefs focusing on the prior information.

To illustrate how our estimates are affected as our beliefs change,  
 suppose further that we have a reason to believe
that the disclosed value of temperature belongs to a
specific reservoir. We quantify this belief
 by assigning a probability with numerical value 
$\gamma$ ($0 \le \gamma \le 1$) to the hypothesis
that the disclosed value $T$ belongs to the initially hot reservoir
(henceforth labeled A). The parameter $\gamma$ is assumed to be
independent of the $T$ value. 

Now we know that {\it if} one of the final temperatures is $T$,
the corresponding value for the other reservoir definitely is $T_+ T_-/T$. 
But as per our  beliefs, the final temperature of reservoir A is: 
(i) $T$, with probability $\gamma$, and (ii)
$T_+ T_-/T$, with probability $1-\gamma$.
Then upon knowing the value $T$, the expected final temperature of reservoir A
 may be defined as a weighted average over the two values:
\be
\overline{T}_A = \gamma T + (1-\gamma) \frac{T_+T_-}{T}.
\label{bart}
\ee
This is our estimate for the final temperature of reservoir A, given 
the information that it is a maximum work process and, one of the final 
temperatures is $T$. Correspondingly, for reservoir B, we should have:
\be 
\overline{T}_B = (1-\gamma) T + \gamma \frac{T_+T_-}{T}.
\label{bbrt}
\ee
Now we use these values to estimate further other
quantities, which are relevant to the performance of the engine.
Our estimate of the heat absorbed by the engine from reservoir A, is
 given by 
\be
Q_{\rm in} = T_+ - \overline{T}_A.
\label{qin}
\ee
Similarly,  our estimate for the heat rejected to reservoir B, will be: 
\be
Q_{\rm out} = \overline{T}_B - T_-.
\label{qout}
\ee
The estimate for work defined as: $W = Q_{\rm in} -Q_{\rm out}$,
turns out to be $W = T_+ +T_- -T  -{T_+ T_-}/{T}$, i.e.
equal to the actual work performed. In particular, the estimate for work
is independent of the parameter $\gamma$, 
showing that the work is not affected
by label uncertainty. 

For brevity, we now calibrate all the temperatures, relative to 
the initial temperature of reservoir A, and  define $\theta = T_-/T_+$.
Then the expected work for a given value $T$,
is 
\be
W = 1+ \theta -T - \frac{\theta}{T}.
\label{wtth}
\ee
Finally, we note that  
the estimate for the efficiency $\eta = W/Q_{\rm in}$, given by 
\be
\eta_\gamma(T) =  \frac{T + \theta T - T^2-\theta}{T-\gamma
T^2-(1-\gamma)\theta},
\label{etg}
\ee
is also affected by the label uncertainty.

Let us now look at the behavior of the above efficiency for 
some special values of $\gamma$.
When $\gamma=1$ ($0$), it corresponds to the case
when we are certain that the disclosed value $T$ belongs to reservoir A (B). 
In these cases, Eq. (\ref{etg}) reduces to $\eta_0 = 1-T$
and $\eta_1 = 1-\theta/T$, respectively.
On the other extreme, when we are maximally uncertain
about the label for the temperature $T$, or
 which means  equal probabilities for $T$ to belong to any of the 
two reservoirs, then we must assign $\gamma=1/2$. Then from Eq. (\ref{etg}), 
the efficiency is expected to be 
\be
\eta_{\frac{1}{2}}(T) = \frac{2(T + \theta T - T^2-\theta)}{(2 T-T^2-\theta)}.
\label{etat}
\ee
Now there are two quantities of interest:

a) {\it Maximum work}: this is obtained by setting
 $\partial W/\partial T = 0$ \cite{Callenbook}, 
which holds at $T = \sqrt{\theta}$.
The efficiency at maximum work is the well-known Curzon-Ahlborn
formula: $\eta_{\rm CA} = 1- \sqrt{\theta}$.

b) {\it Estimated maximum efficiency}: this is given by the condition: 
$\partial \eta_\gamma/\partial T = 0$. From Eq. (\ref{etg}), 
the maximum value is:
\be
\eta^{*}_{\gamma} = \frac{1-\theta}{1+\sqrt{4\gamma(1-\gamma)\theta}}.
\label{etstar}
\ee
\begin{figure}[h]
\includegraphics[height=5.5cm,width=8cm]{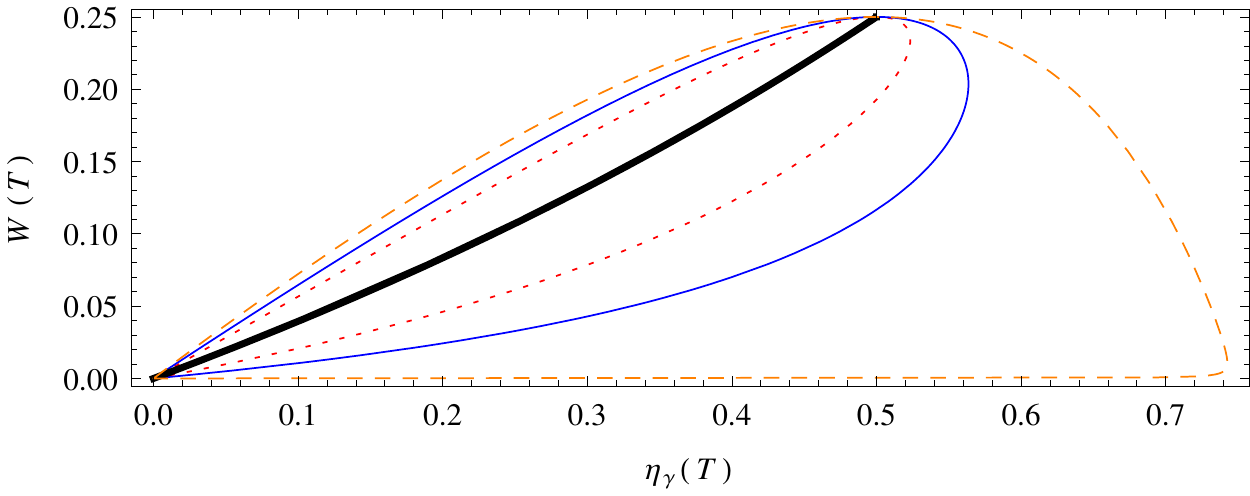}
\caption{A parametric plot between $W(T)$ as in Eq. (\ref{wtth}) and 
efficiency in Eq. (\ref{etg}), for $\theta=0.25$. $T$ takes 
values in the range $[\theta,1]$ and  
$\gamma$ is kept fixed: 1/2 (thick black, monotonic curve), 1/4 (dotted), 1/8 (thin),
 and 1/10000 (dashed). The point for maximum work remains the same, while the 
maximum in the efficiency shifts with $\gamma$. There is a finite work obtained
at the maximum of efficiency, 
which reduces and goes to zero as $\gamma \to 0$ (or $\gamma \to 1$),
whereby the maximum efficiency reaches Carnot value, $1-\theta = 0.75$.}
\label{etfigg}
\end{figure}
From Eq. (\ref{etstar}), we see that in both   
cases of certainty about the labels, implying maximal information,
 the maximum efficiency is the Carnot value. 
But  an uncertainty in the exact labels reduces the 
maximal efficiency ($\eta^{*}_{\gamma}$) below the Carnot value, and 
it reduces to CA value,  in the case of
 maximal uncertainty ($\gamma=1/2$).
Thus the upper bound for efficiency is
related here directly with our state of knowledge and the CA 
efficiency emerges from an entirely different perspective.
Here we have a mechanism which shows how inference under
incomplete information leads us to expect a lower value for the maximum
efficiency obtainable from a thermal engine. 
This feature is further exhibited in 
Fig. \ref{etfigg}, which shows the work versus efficiency curves,
each with a fixed value of $\gamma$. In general, the maximum work and the maximum 
efficiency points are different. At $\gamma=0$ and $\gamma=1$,
they are maximally separate, but tend to merge together 
as $\gamma \to 1/2$.
\begin{figure}
\includegraphics[width=8cm]{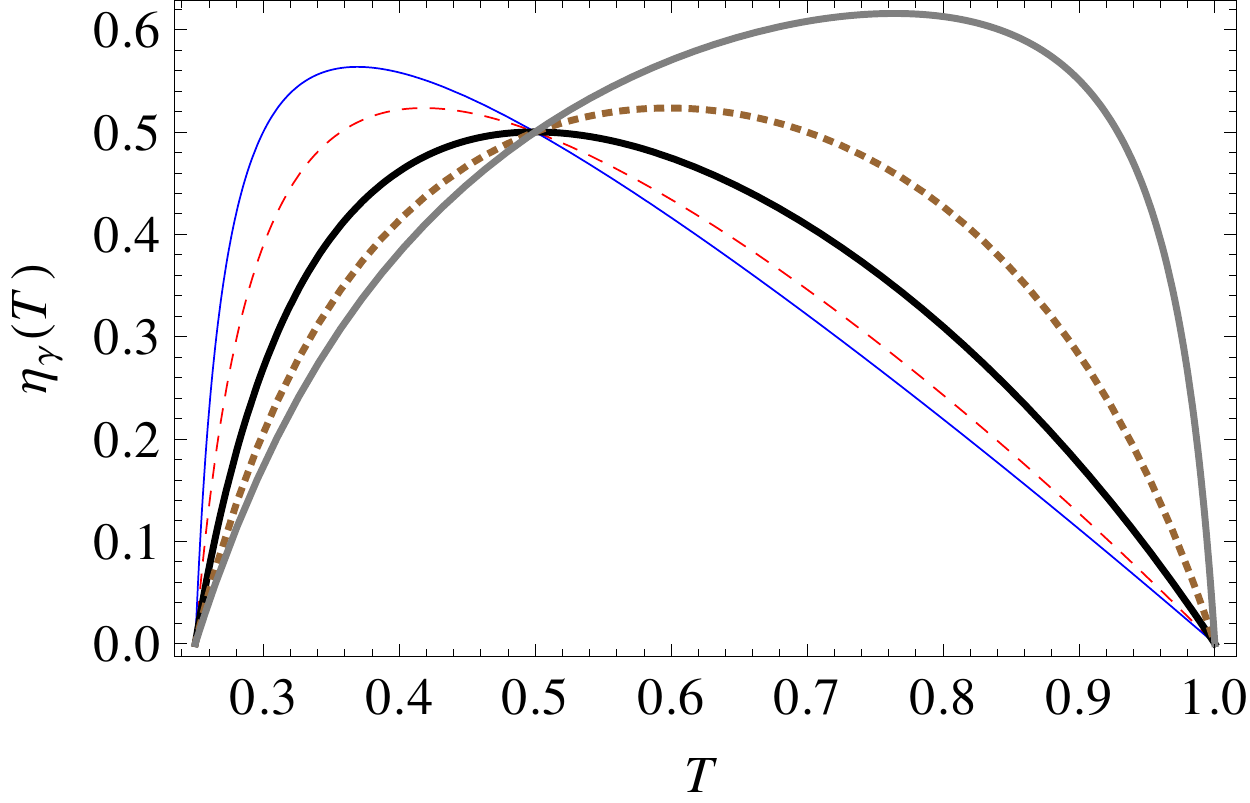}
\caption{Plot of Eq. (\ref{etg}) for $\theta=0.25$. $T$ takes 
values in the range $[\theta,1]$. 
$\gamma$ varies as 1/2 (thick black), 1/4 (dashed), 1/8 (thin),
3/4 (dotted) and 19/20 (thick gray). The maximum value of efficiency
for a given $\gamma$ is given by Eq. (\ref{etstar}). As $\gamma \to 0$
or $\gamma \to 1$, the maximal value approaches the carnot value, $1-\theta =
0.75$.}
\label{etfig2}
\end{figure}
\subsection{Analogy with irreversible model}
The fact that the maximal efficiency drops below the  
Carnot efficiency, indicates that we are infering an 
irreversible behavior  for an otherwise
reversible model with incomplete information.
In this section, 
we suggest an irreversible physical process which 
corresponds to the  above picture obtained through  
inference.

The reservoirs A and B are initially at 
(scaled) temperatures $1$ and $\theta$,
respectively.
Now imagine a two-step process. In the first step, work 
is extracted by coupling reservoirs with a reversible
work source, so that at the end of this step,
 reservoir A reaches temperature $T$ and consequently,
reservoir B is at temperature $\theta/T$.
The work performed is given by Eq. (\ref{wtth}). 
In the second step, the two reservoirs are detached from 
the work source and  put in mutual thermal contact.
Heat is exchanged between them, conserving their
total internal energy. Let at the end of the second-step,
the temperatures of the reservoirs be $\overline{T}_A$
and $\overline{T}_B$, as given by Eqs. (\ref{bart})
and (\ref{bbrt}). The second step is intrinsically
irreversible.  The net heat released by reservoir A is:
$Q_{\rm in} = 1-\overline{T}_A$ and the net heat rejected
to reservoir B is:
$Q_{\rm out}=\overline{T}_B - \theta$. 
These quantities are the same as obtained through inference in Section II.B.

In the above two-step process, we are assigning definite 
temperatures to the reservoirs A and B. Due to the second law,
we expect that overall the heat flows from hot to cold reservoir. 
So at the end of work 
extracting process, we require that 
\be
T\ge \theta/T.
\label{tth}
\ee
If $T=\theta/T$, then the second step is redundant. 
So if $T>\theta/T$ holds then
by the end of second step, we also expect that 
\be
\overline{T}_A \ge \overline{T}_B.
\label{tab}
\ee
It can be easily seen that the above condition requires $\gamma \ge 1/2$,
with $\overline{T}_A = \overline{T}_B$ implying $\gamma=1/2$.
\section{Average estimate of efficiency}
\begin{figure}[h]
\includegraphics[width=8cm]{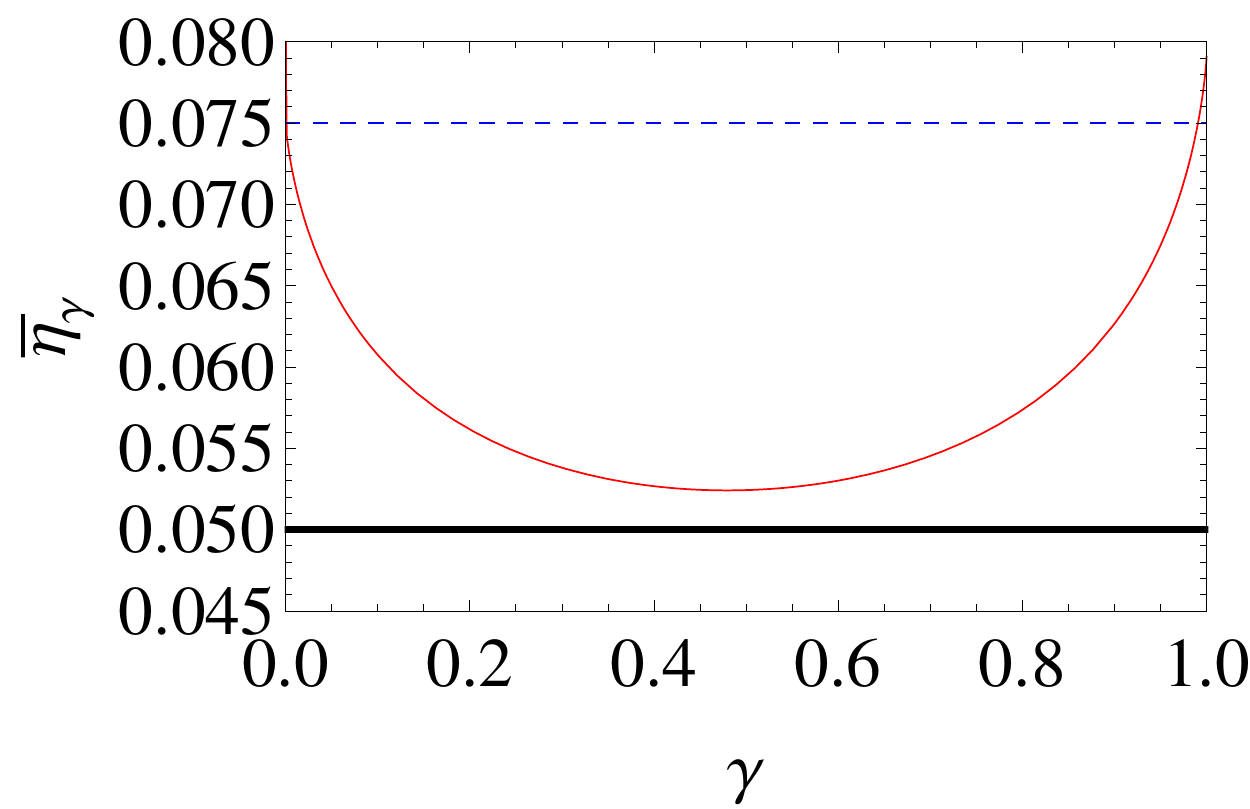}
\caption{Plot of the average efficiency over uniform prior versus $\gamma$,
for $\theta=0.85$. The horizontal lines denote  $(1-\theta)/2$ and  $(1-\theta)/3$
which act as lower bounds for the case $\gamma=1$, and $\gamma=1/2$,
respectively. For $\gamma=0$, the efficiency estimate is exactly $(1-\theta)/2$. }
\label{etca}
\end{figure}

So far, we have assumed that the values of final 
temperatures are pre-specified. 
For instance, Eq. (\ref{etat}) provides an estimate of efficiency 
for a given value $T$ but with complete
ignorance about the reservoir labels. If we are only provided the value $T$,
then we must take $\gamma=1/2$ and our estimate for efficiency
will be Eq. (\ref{etat}).  
In this section, we  extend the game of guessing further 
and estimate the efficiency {\it before} we are provided
the value $T$. So now we assume to be ignorant about
the value $T$ also. To quantify our guess in the absence of
the value $T$, we have to specify a prior 
distribution for $T$ \cite{PRD2012, GPRD2012}, which takes into account
our belief as to which value $T$ from the allowed range, the controller may 
be holding. If we do not have a reason
to expect one value over another, then  all allowed values
are equally likely in the interval
$[T_-,T_+] \equiv [\theta,1]$.
So we must adopt a uniform prior distribution for $T$.
 Our average estimate for efficiency,
defined as the mean value over this uniform prior is then 
\be
\overline{\eta}_{\frac{1}{2}}(\theta) = \int_{\theta}^{1}
 \eta_{\frac{1}{2}} (T) \! \frac{dT}{1-\theta}.
\label{overet}
\ee 
On using Eq. (\ref{etat}) and solving Eq. (\ref{overet}), we obtain 
\bea 
\bar{\eta}_{\frac{1}{2}} &=&  2- \left(1 + \frac{1}{\sqrt{1-\theta}}\right) \ln\left( 1+ 
\sqrt{1-\theta} \right) \nonumber \\
& & -\left(1 - \frac{1}{\sqrt{1-\theta}} \right) \ln\left( 1 -  \sqrt{1-\theta}
\right).
\label{etu}
\eea
For close to equilibrium situations, i.e. $1-\theta \approx 0$,  
the average value of efficiency behaves as follows
\be
\bar{\eta}_{\frac{1}{2}} = \frac{1-\theta}{3}-\frac{(1-\theta)^2}{10} + O[1-\theta]^3.
\ee
On the other extreme, consider the two cases of certainty about the reservoir labels.
For the special case of $\gamma=0$, we have 
$\eta_{0}(T) = 1-T$,
whose average over the uniform prior is
\be 
 \overline{\eta}_{0} = \frac{1-\theta}{2}.
\label{e0m}
 \ee
Also for $\gamma=1$, we have $\eta_{1}(T) = 1-\theta/T$
and the average value over uniform prior, is given by  
\be 
 \overline{\eta}_{1} = 1+ \frac{\theta \ln \theta}{1-\theta},
\label{e1m}
 \ee
 whose expansion behaves as: 
 $\overline{\eta}_{1} \approx (1-\theta)/2 + (1-\theta)^2/6 + O[1-\theta]^3$.
Thus, both of the above averages yield that for near-equilibrium conditions, 
the average efficiency is given 
by $(1-\theta)/2$. This result holds if the reservoir labels are known 
i.e. we know which reservoir temperature 
is chosen as the uncertain variable. In contrast, we obtain  one-third of Carnot value,
 if we are maximally uncertain about the specific reservoir labels.
This is the main result of the paper, that the expected  
efficiency near equilibrium falls under two different classes,
 determined by the state of our knowledge about the system.
Fig. \ref{etca} shows this dependence on $\gamma$ in a more
clear fashion. 
\section{Summary}
Inference is a kind of common sense reasoning
in the face of incomplete information. It seeks to provide the most
unbiased guess consistent with the constraints and laws
obeyed by the system. We have considered  
a plausible use of this
reasoning for classical thermodynamic machines operating 
under reversible conditions. An important new kind of uncertainty
studied in this paper relates to incomplete knowledge about the 
labels of the uncertain parameters. We define a guessing game which
combines features of subjective ignorance with the objective
matters of fact about the physical system.  
Some interesting observations made are that the infered
behavior has features of irreversibility, for example, the 
maximum efficiency of the engine is below Carnot limit. We also 
suggested an analog physical model which mimics the  
estimates of thermodynamic quantities in the infered model.
It is not known if the mapping to an actual
physical model is always guaranteed from a given inference based 
model. Further, we cannot be sure about the uniqueness of this
mapping, and there may exist more than one physical 
processes which simulate the consequences of our game,
or conversely, more than one games which simulate
a physical process.
However, in our opinion, the possibility of a mapping to 
an objective physical model indicates that inference
 anticipates a behavior which is allowed by the physical laws. Thus 
the mapping also reassures that the estimates are consistent
with the laws of thermodynamics.

We have estimated an average efficiency based on the 
prior probability distribution and found that the so called label uncertainty
makes the efficiency drop to 1/3 of Carnot value in near-equilibrium
regime. On the other hand, if there is no label uncertainty,
the infered efficiency is equal to 1/2 of Carnot value. 
This is consistent with our expectation that more information
we lack, less is the efficiency we expect. Still the goal of
inference is not to predict  the ``true'' behavior of
physical models proper, but to make a rational guess 
based on incomplete information. 

In this paper we have defined a set of specific ``games'' representing additional 
subjective lack of information.
This setting supplements the ``objective ignorance'' present already due to the 
laws of thermodynamics. Nevertheless, it turns out that the combined model may 
be equivalent to a ``purely physical'' model with some specific irreversible 
features -- a remarkable non-uniqueness in terms of interpretation.
It is hoped that the approach and conclusions
of this paper may help to further develop procedures
which incorporate different kinds of prior information,
which then could shed light on the subtle interplay
between subjective and objective ignorance
presumed in the modeling of natural phenomena like, e.g., in the
context of (rather complex) climate models.

\end{document}